\documentstyle[12pt]{article} 
\newcommand{\swav}[1]{ \stackrel{\sim}{#1} }

\newcommand{\ehat}{ \hat U_{\epsilon} } 
 
\newcommand{\define}{ \stackrel{\triangle}{=} } 
 
\def\be{\begin{equation}} 
\def\ee{\end{equation}} 
\def\ba{\begin{array}} 
\def\ea{\end{array}} 
\begin{document} 
\title{\bf Differential Geometrical Formulation 
    of Gauge Theory of Gravity } 
\author{{Ning Wu}$^1$ 
\thanks{email address: wuning@mail.ihep.ac.cn}, 
{\rm Zhan Xu}$^2$, {\rm Dahua Zhang}$^1$ 
\\ 
\\ 
{\small Institute of High Energy Physics, P.O.Box 918-1, 
Beijing 100039, P.R.China}$^1$ \\ 
\\ 
{\small Tsinghua University, Beijing 100080, P.R.China }$^2$ 
} 
\maketitle 
\vskip 0.5in 
 
~~\\ 
PACS Numbers: 11.15.-q, 04.60.-m, 02.40.-k. \\ 
Keywords: gauge field, quantum gravity, geometry.\\

\vskip 0.5in 
 
\begin{abstract} 
 
Differential geometric formulation of quantum gauge theory 
of gravity is studied in this paper. The quantum gauge theory 
of gravity which is proposed in the references hep-th/0109145 
and hep-th/0112062 
is formulated completely in the framework of traditional  
quantum field theory. In order to study the relationship 
between quantum gauge theory of gravity and traditional 
quantum gravity which is formulated in curved space, it 
is important to find the differential geometric formulation  
of quantum gauge theory of gravity. We first give out the 
correspondence between quantum gauge theory of gravity and 
differential geometry. Then we give out differential geometric 
formulation of quantum gauge theory of gravity. 
\\ 
 
\end{abstract} 
 
\newpage 
 
\Roman{section} 
 
\section{Introduction} 
\setcounter{equation}{0} 
 
It is known that, in classical Newton's theory of gravity, 
gravity is treated as  physical interactions between two 
massive objects, and gravity does not affect the structure 
of space-time\cite{01}. In Einstein's general theory of 
relativity, gravity is treated as geometry of  
space-time\cite{02,03}. In other words, in general relativity, 
gravity is not treated as a physical interactions, it is 
a part of structure of space-time. Inspired by Einstein's 
general theory of relativity, traditional relativistic theory 
of gravity and traditional canonical quantum theory of 
gravity is formulated in the framework of differential  
geometry\cite{b1,b2,b3,b4,b5,b6}. \\ 
 
Recently, based on completely new notions and completely  
new methods, Wu proposes a new quantum gauge theory of  
gravity, which is based on gauge principle\cite{1,3}. 
The central idea is to use traditional gauge field 
theory to formulate quantum theory of gravity. 
This new quantum gauge theory of gravity is 
renormalizable\cite{1,2}. A strict formal proof on  
the renormalizability of the theory is also given in  
the reference \cite{1,2}. This new quantum gauge theory 
of gravity is formulated in the flat 
physical space-time, which is 
completely different from that of the traditional  
quantum gravity at first appearance. But this difference 
is not essential, for quantum gauge theory of gravity 
can also be formulated in curved space.  
In gauge theory of gravity, gravity is treated as 
physical interactions, but in this paper, gravity 
is treated as space-time geometry. It means that gravity 
has physics-geometry duality, which is the nature 
of gravitational interactions. In this paper, we first 
give out the correspondence between quantum gauge 
theory of gravity and differential geometry. Then, 
we give out the differential geometrical formulation 
of quantum gauge theory of gravity.\\

\section{Correspondence Between Quantum Gauge Theory 
    of Gravity and Differential Geometry} 
\setcounter{equation}{0} 
 
In gravitational gauge theory, there are two different kinds 
of space-time: one is the traditional Minkowski space-time which 
is the physical space-time, another is gravitational gauge group 
space-time which is not a physical space-time. Two space-times have 
different physical meanings. In differential geometry, there are also 
two different space-times. This can be seen from the tetrad 
field in Cartan form. In cartan form, there are two different  
space-times, one is base manifold, another is the tangent space 
in Cartan tetrad. The correspondence between these two 
different spaces in two different theories are: the traditional 
Minkowski space-time in gravitational gauge theory corresponds to 
the tangent space in Cartan tetrad; the gravitational  
gauge group space-time corresponds to the base manifold of 
differential geometry. \\

In gravitational gauge theory, the metric in Minkowski space-time 
is the flat metric $\eta^{\mu\nu}$, which corresponds to the flat 
metric $\eta^{ab}$ in Cartan tetrad. The metric in  
gravitational gauge group space-time is the curved metric  
$g^{\alpha \beta}$ which corresponds to the curved metric  
$g^{\mu\nu}$ in differential geometry. \\ 
 
In gravitational gauge theory, $G^{\alpha}_{\mu}$ is defined by 
\be \label{2.1}  
G_{\mu}^{\alpha} = \delta_{\mu}^{\alpha} - g C_{\mu}^{\alpha}, 
\ee 
which corresponds to the Cartan tetrad field 
$e_{a.}^{~\mu}$. $G^{-1 \mu}_{\alpha}$ corresponds to reference 
frame field $e^a_{. \mu}$. That is
\be \label{2.101}
G_{\mu}^{\alpha} \Longleftrightarrow  e_{a.}^{~\mu},
\ee
\be \label{2.101}
G^{-1\mu}_{\alpha} \Longleftrightarrow  e^{a}_{.~\mu}.
\ee
Therefore, the following two  
relations have the same meaning, 
\be \label{2.3}  
g^{\alpha\beta} = \eta^{\mu\nu}  
G_{\mu}^{\alpha} G_{\nu}^{\beta} 
 \Longleftrightarrow  
g^{\mu\nu} = \eta^{a b}  
e_{a .}^{~\mu} e_{b .}^{~\nu}, 
\ee 
where $g^{\alpha\beta}$ is the metric in gravitational gauge  
group space-time, and $g^{\mu\nu}$ is the metric of the base manifold 
in differential geometry. Similarly, the following two relations  
correspond to each other, 
\be \label{2.4}  
g_{\alpha\beta} = \eta_{\mu\nu}  
G^{-1\mu}_{\alpha} G^{-1\nu}_{\beta} 
 \Longleftrightarrow 
g_{\mu\nu} = \eta_{a b}  
e^{a }_{.~ \mu} e^{b }_{. ~\nu}. 
\ee 
\\ 
 
Define  
\be \label{2.6}  
\partial_a \define e_{a .}^{~ \mu} \partial_{\mu}, 
\ee 
which is the derivative in Cartan tetrad.  
It corresponds to the gravitational gauge covariant derivative 
$D_{\mu}$ in gravitational gauge theory
\be \label{2.601}
D_{\mu} \Longleftrightarrow
\partial_a.
\ee 
\\

In gravitational gauge theory, all matter field, such as  
scalar fields $\phi(x)$, Dirac field $\psi (x)$, vector field 
$A_{\mu}(x)$, $\cdots$, are fields in Minkowski space, which  
correspond to fields in Cartan tetrad. In other words, 
all matter fields in differential geometry are defined in  
Cartan tetrad. \\

Finally, what is the transformation in differential geometry 
which corresponds to the the gravitational gauge transformation?  
Under most general coordinate transformation, the transformation 
of Cartan tetrad field is : 
\be \label{2.7}  
e_{a .}^{~\mu}  \to e^{\prime ~ \mu}_{a .} 
= \frac{\partial x^{\prime \mu}}{\partial x^{\nu}} 
L_{a .}^{~b} e^{~\nu}_{b .}, 
\ee 
where $L_{a .}^{~b}$ is the associated Lorentz transformation.  
As we have stated before, under gravitational gauge transformations, 
there is no associated Lorentz transformation. In other words,  
under gravitational gauge transformation,  
\be \label{2.8} 
L_{a .}^{~b} = \delta_a^b. 
\ee 
Therefore, the gravitational gauge transformation of 
Cartan tetrad field is 
\be \label{2.9}  
e_{a .}^{~\mu}  \to e^{\prime ~ \mu}_{a .} 
= \frac{\partial x^{\prime \mu}}{\partial x^{\nu}} 
 e^{~\nu}_{a .}, 
\ee 
We call this transformation translation transformation. So, gravitational
gauge transformation in gravitational gauge theory corresponds to the
translation transformation in differential geometry. 
According to eq.(\ref{2.6}),  $\partial_a$ does not change under 
translation transformation, 
\be \label{2.10}  
\partial_a \to \partial_a^{\prime} = \partial_a. 
\ee 
Eq.(\ref{2.9}) corresponds to the following transformation in  
gravitational gauge theory, 
\be \label{2.11}  
G_{\mu}^{\alpha} \to G_{\mu}^{\prime \alpha} 
= \Lambda^{\alpha}_{~\beta} (\ehat G_{\mu}^{\beta} ). 
\ee 
Eq.(\ref{2.10}) corresponds to the following gravitational gauge  
transformation in gravitational gauge theory, 
\be \label{2.12}  
D_{\mu} \to D'_{\mu} 
= \ehat D_{\mu} \ehat^{-1}. 
\ee 
\\

Finally, as a summary, we list some important correspondences
between physical picture and geometry picture 
in the following table 1.

\begin{table}[htp]
\begin{center}
\doublerulesep 0pt
\renewcommand\arraystretch{1.5}
\begin{tabular}{|l|l|}
\hline
 
Quantum Gauge Theory of Gravity &  Differential Geometry  \\
\hline 
\hline  

gauge group space-time   &  base manifold  \\

Minkowski space-time     &  tangent space in tetrad  \\
 
index $\alpha$           &  index  $\mu$   \\

index  $\mu$             &  index  $a$    \\
\hline

$G_{\mu}^{\alpha}$       & $ e_{a.}^{~\mu}$  \\

$G^{-1\mu}_{\alpha}$     & $e^{a}_{.~\mu}$   \\

$g^{\alpha\beta} = \eta^{\mu\nu}
G_{\mu}^{\alpha} G_{\nu}^{\beta}  $     &

$g^{\mu\nu} = \eta^{a b}
e_{a .}^{~\mu} e_{b .}^{~\nu}$    \\

$g_{\alpha\beta} = \eta_{\mu\nu}
G^{-1\mu}_{\alpha} G^{-1\nu}_{\beta} $    &
$g_{\mu\nu} = \eta_{a b}
e^{a }_{.~ \mu} e^{b }_{. ~\nu}$     \\
\hline

$D_{\mu}$               &   $\partial_a$   \\

$F_{\mu\nu}^{\alpha}$   &   $\bar \omega^{\mu}_{a b}$   \\
\hline 

gravitational gauge transformation  &  translation transformation  \\

gravitational gauge covariant       &  translation invariant   \\

$\Lambda^{\alpha}_{~\beta}$  &
$\frac{\partial x^{\prime \mu}}{\partial x^{\nu}}$   \\

$\Lambda_{\alpha}^{~\beta}$  &
$\frac{\partial x^{\nu}}{\partial x^{\prime \mu}}$   \\

$G_{\mu}^{\alpha} \to G_{\mu}^{\prime \alpha}
= \Lambda^{\alpha}_{~\beta} (\ehat G_{\mu}^{\beta} )$   &
$e_{a .}^{~\mu}  \to e^{\prime ~ \mu}_{a .}
= \frac{\partial x^{\prime \mu}}{\partial x^{\nu}}
 e^{~\nu}_{a .}$    \\

$G^{-1 \mu}_{\alpha} \to G^{\prime -1 \mu}_{ \alpha}
= \Lambda_{\alpha}^{~\beta} (\ehat G^{-1\mu}_{\beta} )$   &
$e^{a}_{.~\mu}  \to e_{. ~ \mu}^{\prime a}
= \frac{\partial x^{\nu}}{\partial x^{\prime \mu}}
 e_{.~\nu}^{a}$    \\

$g^{\alpha\beta} \to g^{\prime \alpha\beta}
= \Lambda^{\alpha}_{~\alpha_1} \Lambda^{\beta}_{~\beta_1}
(\ehat g^{\alpha_1 \beta_1} )$   &
$g^{\mu\nu}  \to g^{\prime \mu\nu}
= \frac{\partial x^{\prime \mu}}{\partial x^{\mu_1}}
\frac{\partial x^{\prime \nu}}{\partial x^{\nu_1}}
 g^{\mu_1 \nu_1}$    \\

$g_{\alpha\beta} \to g^{\prime}_{ \alpha\beta}
= \Lambda_{\alpha}^{~\alpha_1} \Lambda_{\beta}^{~\beta_1}
(\ehat g_{\alpha_1 \beta_1} )$   &
$g_{\mu\nu}  \to g^{\prime}_{ \mu\nu}
= \frac{\partial x^{\mu_1}}{\partial x^{\prime \mu}} 
\frac{\partial x^{\nu_1}}{\partial x^{\prime \nu}}                            
 g_{\mu_1 \nu_1}$    \\

$D_{\mu} \to D'_{\mu}
= \ehat D_{\mu} \ehat^{-1}$    &
$\partial_a \to \partial_a^{\prime} = \partial_a$   \\

$F_{\mu\nu}^{\alpha} \to F_{\mu\nu}^{\prime \alpha}
=\Lambda^{\alpha}_{~\beta} (\ehat F_{\mu\nu}^{\beta} )$&
$\bar \omega^{\mu}_{a b} \to
\bar \omega^{\prime \mu}_{a b}
= \frac{\partial x^{\prime \mu}}{\partial x^{\nu}}
\bar \omega^{\nu}_{a b}$                            \\

\hline
\end {tabular}
\caption { Correspondence between two pictures of gravity. }
\end{center}
\end{table}

\section{ Differential Geometrical Formulation of
    Gravitational Gauge Theory} 
\setcounter{equation}{0} 

In differential geometrical formulation of gravitational gauge
theory, all fields are expressed in Cartan orthogonal tetrad.
In differential geometry, gravitational field is represented
by tetrad field $e_{a .}^{~ \mu}$. The field strength of
gravitational field is denoted by $\bar \omega^{\mu}_{a b}$,
\be \label{3.1}  
\bar \omega^{\mu}_{a b} = - \frac{1}{g}
\left\lbrack
(\partial_a e^{~\mu}_{b .}) 
- (\partial_b e^{~\mu}_{a .} )
\right\rbrack. 
\ee 
$\bar \omega^{\mu}_{a b}$ corresponds to the field strength tensor
$F^{\alpha}_{\mu\nu}$ in gravitational gauge theory. Under 
translation transformation, $\bar \omega^{\mu}_{a b}$
transforms as
\be \label{3.2}  
\bar \omega^{\mu}_{a b} \to
\bar \omega^{\prime \mu}_{a b} 
= \frac{\partial x^{\prime \mu}}{\partial x^{\nu}}
\bar \omega^{\nu}_{a b}. 
\ee 
\\

It is known that, in differential geometry, covariant
derivative of cartan tetrad vanishes
\be \label{3.201}
D_{\mu} e^a_{.~\nu}=0.
\ee
It gives out the following relation
\be \label{3.202}
\swav{\Gamma}^a_{bc} 
= e^a_{.~\lambda} e_{b .}^{~\nu} e_{c .}^{~\mu}
\Gamma^{\lambda}_{\mu\nu}
+ e^a_{.~\nu} (\partial_{\mu} e_{b.}^{~ \nu} )
e^{~\mu}_{c.}, 
\ee
where $\Gamma^{\lambda}_{\mu\nu}$ is the affine connexion
and $\swav{\Gamma}^a_{bc}$ is the Cartan connexion. 
From this relation, we can obtain
\be \label{3.3}  
\swav{T}^a_{bc} 
= e^a_{.~\lambda} e_{b~.}^{~\mu} e_{c~.}^{~\nu}
T^{\lambda}_{\mu\nu}
+ g e^a_{.~\mu} \bar \omega^{\mu}_{bc}
\ee 
where $T^{\lambda}_{\mu\nu}$ and $\swav{T}^a_{bc}$ are
torsion tensors
\be \label{3.301}
\swav{T}^a_{bc}
= \swav{\Gamma}^a_{bc} - \swav{\Gamma}^a_{cb},
\ee
\be \label{3.302}
T^{\lambda}_{\mu\nu}
= \Gamma^{\lambda}_{\mu\nu} - \Gamma^{\lambda}_{\nu\mu}.
\ee
For gravitational gauge theory, the affine connexion 
is the Christoffel connexion. For  Christoffel connexion,
the torsion $T^{\lambda}_{\mu\nu}$ vanish. Then
\be \label{3.4}
\swav{T}^a_{bc} 
= g e^a_{.~\mu} \bar \omega^{\mu}_{bc}.
\ee 
It seems that the field strength of gravitational field is
related to the torsion of Cartan connexion. 
\\

Translation transformation of metric tensor $g_{\mu\nu}$ is
\be \label{3.6}  
g_{\mu\nu} \to
g'_{\mu\nu} 
= \frac{\partial x^{\mu_1}}{\partial x^{\prime \mu}}
\frac{\partial x^{ \nu_1}}{\partial x^{\prime \nu}}
g_{\mu_1 \nu_1}. 
\ee 
The metric in Cartan tetrad is denoted as $\eta^{a b}$,
which is the traditional Minkowski metric. Under translation
transformation, $\eta^{a b}$ is invariant
\be \label{3.8}  
\eta^{a b} \to \eta^{\prime a b} = \eta^{ab}.
\ee 
\\

The lagrangian ${\cal L}$ for pure gravitational field is selected as
\be \label{3.9}  
{\cal L} = - \frac{1}{4}
\eta^{ac} \eta^{bd} g_{\mu\nu}
 \bar \omega^{\mu}_{a b}
 \bar \omega^{\nu}_{c d}.
\ee 
Using eq.(\ref{2.4}) and eq.(\ref{3.4}), we can change the above 
lagrangian density into the following form
\be \label{3.901}
{\cal L} = - \frac{1}{4 g^2}
\eta^{ac} \eta^{bd} \eta_{ef}
\swav{T}^e_{ ab } \swav{T}^f_{ cd }.
\ee
It is easy to prove that this lagrangian is invariant under 
translation transformation. The concept of translation 
invariant in differential  geometry corresponds to the 
concept of gravitational gauge covariant in 
gravitational gauge theory. 
\\

In order to introduce translation invariant action of the system, we need
to introduce a factor which is denoted as $J(C)$ in gravitational gauge
theory. In this paper, $J(C)$ is selected as
\be \label{3.11}  
J(C) = \sqrt{ - {\rm det} (g_{\mu\nu}) }.
\ee 
The action of the system is defined by
\be \label{3.12}  
S = \int {\rm d}^4x ~J(C) ~{\cal L}.
\ee 
This action is invariant under translation transformation. \\

For real scalar field $\phi$, its gravitational interactions are
described by
\be \label{3.13}  
{\cal L} = - \frac{1}{2} (\partial_a \phi) (\partial_b \phi)
- \frac{1}{2} m^2 \phi^2, 
\ee 
where $m$ is the mass of scalar. For complex scalar field, its lagrangian
is selected to be
\be \label{3.14}  
{\cal L} = -  (\partial_a \phi)^{\ast} (\partial_b \phi)
- m^2 \phi^{\ast} \phi. 
\ee 
Under translation transformations, $\phi(x)$ and $\partial_a \phi(x)$
transform as
\be \label{3.15}
\phi (x) \to \phi'(x') = \phi(x),
\ee 
\be \label{3.16}
\partial_a \phi (x) \to \partial'_a \phi'(x') 
= \partial_a \phi(x).                            
\ee 
Using eq.(\ref{3.15}) and eq.(\ref{3.16}), we can prove that the
lagrangians which are given by eq.(\ref{3.13}) and eq.(\ref{3.14})
are translation invariant. \\

For Dirac field, its lagrangian is given by 
\be \label{3.17}
{\cal L} = - \bar \psi (\gamma^a \partial_a + m ) \psi.
\ee
Under translation transformations,
\be \label{3.18}
\psi (x) \to \psi'(x') = \psi(x),
\ee
\be \label{3.19}
\partial_a \psi (x) \to \partial'_a \psi'(x') 
= \partial_a \psi(x),
\ee
\be \label{3.20}
\gamma^a \to \gamma^{\prime a}  = \gamma^a.
\ee
Under these transformations, the lagrangian given by
eq.(\ref{3.17}) is invariant. \\

For vector field, the lagrangian is given by
\be \label{3.21}
{\cal L} = - \frac{1}{4} \eta^{ac} \eta^{bd}
A_{ab} A_{cd} 
- \frac{m^2}{2} \eta^{ab} A_a A_b,
\ee 
where 
\be \label{3.22}
A_{ab} = \partial_a A_b - \partial_b A_a,
\ee 
which is the field strength of vector field. This lagrangian
is invariant under the following translation transformations
\be \label{3.23}
A_a (x) \to A'_a (x') = A_a (x),
\ee 
\be \label{3.24}
A_{ab} (x) \to A'_{ab} (x') = A_{ab} (x),
\ee 
\be \label{3.25}
\eta^{ab} \to \eta^{\prime ab} = \eta^{ab}.
\ee 
\\

For $U(1)$ gauge field, its lagrangian is\cite{1,2,7}
\be \label{3.26}
{\cal L} = - \bar \psi \left\lbrack
\gamma^a (\partial_a -ie A_a) + m \right\rbrack \psi
- \frac{1}{4} \eta^{ac} \eta^{bd}
{\mathbf A}_{ab} {\mathbf A}_{cd},
\ee 
where
\be \label{3.27}  
{\mathbf A}_{ab} =  A_{ab} 
+ g e^c_{.~\mu} A_c \bar\omega^{\mu}_{ab},
\ee 
\be \label{3.28}  
A_{ab} = \partial_a A_b - \partial_b A_a,   
\ee 
In eq.(\ref{3.26}), $e$ is the coupling constant for
$U(1)$ gauge interactions. 
This lagrangian is invariant under the following 
translation  transformation,
\be \label{3.29}
A_a (x) \to A'_a (x') = A_a (x),
\ee 
\be \label{3.30}
{\mathbf A}_{ab} (x) \to {\mathbf A}'_{ab} (x') 
= {\mathbf A}_{ab} (x).
\ee 
It is also invariant under local $U(1)$ gauge 
transformation\cite{1,2,7}.
\\

For $SU(N)$ non-Abel gaueg field, its lagrangian is\cite{1,2,7}
\be \label{3.31}
{\cal L} = - \bar \psi \left\lbrack
\gamma^a (\partial_a -i g_c A_a) + m \right\rbrack \psi
- \frac{1}{4} \eta^{ac} \eta^{bd}   
{\mathbf A}^i_{ab} {\mathbf A}^i_{cd},
\ee 
where
\be \label{3.27}
{\mathbf A}^i_{ab} =  A^i_{ab} 
+ g e^c_{.~\mu} A^i_c \bar\omega^{\mu}_{ab},
\ee 
\be \label{3.28}  
A^i_{ab} = \partial_a A_b - \partial_b A_a
+ g_c f_{ijk} A_a^j A_b^k.
\ee 
In above relations, $g_c$ is the coupling constant for 
$SU(N)$ non-Able gauge interactions. It can be proved
the this lagrangian is invariant under 
$SU(N)$ gauge transformation and translation
transformation\cite{1,2,3,7}. \\

\section{Summary} 
\setcounter{equation}{0}

In this paper, the geometrical formulation of gauge theory
of gravity is studied, which is performed in the
geometrical formulation of gravity\cite{1,3}.  In this 
picture, we can see that gravitational field is put into
the structure of space-time and there is no physical
gravitational interactions in space-time. \\

In gravitational gauge theory, all matter field, such as
scalar fields $\phi(x)$, Dirac field $\psi (x)$, vector field
$A_{\mu}(x)$, $\cdots$, are fields in Minkowski space, which
correspond to fields in Cartan tetrad. In other words,
all matter fields in differential geometry are defined in
Cartan tetrad. \\

In gravitational gauge theory, the symmetry transformation
is gravitational gauge transformation, while in 
differential  geometry, the corresponding
symmetry transformation is
translation transformation. The concept of translation
invariant in differential  geometry corresponds to the
concept of gravitational gauge covariant in
gravitational gauge theory.  \\

\end{document}